\documentstyle[aps,preprint]{revtex}
\newlength{\defaultparindent}
\newenvironment{Default Paragraph Font}{}{}

\begin{document}
\draft
\title{Measurement of the Far Infrared Magneto-Conductivity Tensor of
Superconducting YBa$_2$Cu$_3$O$_{7-\delta }$ Thin Films}
\author{H.-T. S. Lihn, S. Wu, S. Kaplan, H. D. Drew}
\address{Department of Physics, University of Maryland, College Park, MD 20740\\
and\\
Laboratory for Physical Sciences, College Park, MD 20740}
\author{Qi Li\cite{address} and D. B. Fenner}
\address{Advanced Fuel Research, East Hartford, Connecticut 06138}
\date{\today }
\maketitle

\begin{abstract}
We report measurements of the far infrared transmission of superconducting
YBa$_2$Cu$_3$O$_{7-\delta }$ thin films from 5 cm$^{-1}$ to 200 cm$^{-1}$ in
fields up to 14$T$. A Kramers-Kronig analysis of the magneto-transmission
spectrum yields the magneto-conductivity tensor. The result shows that the
magneto-conductivity of YBa$_2$Cu$_3$O$_{7-\delta }$ is dominated by three
terms: a London term, a low frequency Lorentzian ($\omega _1\approx $ 3 cm$%
^{-1}$) of width $\Gamma _1=$ 10 cm$^{-1}$ and a finite frequency Lorentzian
of width $\Gamma _2=$ 17 cm$^{-1}$ at $\omega _2=$ 24 cm$^{-1}$ in the hole
cyclotron resonance active mode of circular polarization.\\
\end{abstract}


\narrowtext

The electrodynamic response of type II superconductors in the mixed state is
strongly affected by vortex dynamics as extensive dc transport and microwave
($\mu $wave) frequency studies have shown for both conventional and high
temperature superconductors\cite{GR,Golo}. For a pinned vortex lattice the
results of the $\mu $wave experiments can be described in terms of a
conductivity function consisting of two terms: a London term reduced from
its zero field strength and a zero frequency oscillator with a width
characterized by the ``depinning frequency'' $\omega _d=\kappa /\eta $ where 
$\kappa $ is the pinning force constant and $\eta $ is the viscosity\cite
{GR,CC}. That this oscillator occurs at zero frequency has been attributed
to the massless dynamics of the vortices. On the other hand a far richer
phenomenology may be anticipated for the electrodynamics of the vortex
system. The possibilities include: a pinning resonance associated with the
inertial motion of the vortex in its pinning field\cite{Hsu}, resonant
excitation of quasiparticles to the quantized levels in the vortex core\cite
{Zhu:Drew,Janko} and, in the clean limit, the collective cyclotron resonance
of the electron system\cite{Karrai,Drew}. Moreover, quite generally, the
response of an electron system in the presence of a magnetic field is
expected to be chiral, so that $\sigma _{xy}\left( \omega ,H\right) \neq 0$%
\cite{Spielman}. Although these effects have not been observed at $\mu $wave
frequencies, they have been reported from recent experiments in high $T_c$
films at far infrared (FIR) frequencies\cite{Karrai,Lihn,Choi}. These
experiments were compared with a clean limit theory of vortex dynamics that
predicts that these three resonances are hybridized leading to two finite
frequency chiral resonances\cite{Choi}. However, this theory does not
predict the zero frequency resonance observed at $\mu $wave frequencies. The
earlier FIR experiments were limited to $\omega >$ 25 cm$^{-1}$ so that they
did not observe the zero frequency oscillator nor did they completely
resolve the pinning resonance\cite{Vortex}. Therefore there is a gap in our
understanding of the electrodynamics of the vortex system that coincides
with the frequency gap in the measurements between the $\mu $wave and FIR.

In this letter we present the magneto-transmission spectrum of YBa$_2$Cu$_3$O%
$_{7-\delta }$ (YBCO) films over the frequency range from 5.26 cm$^{-1}$ to
200 cm$^{-1}$ by using the combination of broad band Fourier Transform
Spectroscopy (FTS) and a FIR laser source. These measurements cover the
entire range of frequencies relevant to the vortex system and provide a
reconciliation of the $\mu $wave and FIR phenomenologies. A Kramers-Kronig
transformation (KKT) technique\cite{Landau} is used to obtain the real and
imaginary parts of the magneto-conductivity tensor as a function of
frequency. The results show that the electrodynamic response of the pinned
vortex system is dominated by two major Lorentzian oscillators in addition
to a London term. These results, together with evidence for the weaker
vortex core resonances reported recently\cite{Lihn}, suggest a vortex
response that combines the features of both the massless vortex model of
Gittleman-Rosenblum (G-R) and the clean limit model of Hsu which includes
the vortex core structure and inertia\cite{GR,Hsu}.

The samples are YBCO thin films grown by pulsed laser deposition on silicon
substrates with yittria stabilized zirconia buffer layers and cap layers.
The film thickness is typically $d=$ 400\AA\ and typical critical
temperatures are $T_c$ = 89$\pm 1$ K measured by ac susceptibility. Their
growth and characterization is described in detail elsewhere\cite{Lihn,Fork}%
. The broadband transmission of the films was measured from 30 cm$^{-1}$ to
200 cm$^{-1}$, using FTS with a 2.2 K bolometer detector. Magnetic fields up
to 14$T$ were applied perpendicular to the a-b plane of the YBCO thin films.
The incident FIR radiation was elliptically polarized by a polarizer
comprised of a metal grid linear polarizer and a x-cut quartz waveplate. The
elliptically polarized transmission data was unfolded, using a calibration
of the polarizer efficiency, to give the circularly polarized response, $%
T^{\pm }(\omega ,H)$. For the low frequency measurements we use a CO$_2$
pumped FIR laser which has useful discrete FIR lines from 5.26 cm$^{-1}$ to
96 cm$^{-1}$. A more detailed description of the experiment and data
manipulation are given elsewhere\cite{Karrai,Choi,Vortex}. Leakage of
radiation around or through the sample is a potential problem at low
frequencies where the transmission of the film is small (as low as 0.02\% at
5 cm$^{-1}$). Leakage was eliminated by the use of Ecosorb and graphite to
absorb stray light\cite{Wu}. The nearly quadratic $T(\omega ,0)$ observed
down to our lowest frequencies attests to the low levels of radiation
leakage in these experiments.

The experiments measured both the absolute transmission at zero field $%
T(\omega ,0)$ and the transmission ratio $T^{\pm }(\omega ,H)/T(\omega ,0)$.
The transmission ratio measurement is more accurate than the absolute
transmission measurement since the sample is undisturbed, which makes it
possible to measure the change induced by vortex dynamics precisely. The
transmission coefficient $T^{\pm }(\omega ,H)=\left| t^{\pm }(\omega
,H)\right| ^2$ where the transmission amplitude $t^{\pm }(\omega ,H)$ is
related to the conductivity by 
\begin{equation}
t^{\pm }(\omega ,H)=(n_{Si}+1)\left/ \left( Z_0d\sigma ^{\pm }(\omega
,H)+n_{Si}+1\right) \right.  \label{transmission}
\end{equation}
in which $Z_0$ is the impedance of free space, $n_{Si}$ is the refractive
index of the silicon substrate (nearly constant and real) and $\sigma ^{\pm
}(\omega ,H)$ is the conductivity in the two circular polarization modes. In
order to eliminate multiple reflection effects in the substrate, we either
averaged the spectrum with a resolution (4 cm$^{-1}$) lower than the spacing
of the interference fringes (1 cm$^{-1}$) in the broadband measurement or
use an anti-reflection coating on the back side of the substrate in the
laser experiment\cite{Wu}.

Fig.\ref{t9} shows $\left| t^{\pm }(\omega ,H)/t(\omega ,0)\right| $ as a
function of frequency $\omega $ at $H$=9$T$ and 4 K. $\left| t(\omega
,0)\right| $ at 4 K is also shown in the inset. The response $\left|
t^{+}(H)/t(0)\right| $ in the electron cyclotron resonance active
polarization (eCP) mode is plotted for positive frequencies and the hole
active polarization (hCP) response $\left| t^{-}(H)/t(0)\right| $ is plotted
for negative frequencies. This representation will be explained below (See
Eq.(\ref{circmap}).). The transmission ratios show a sharp rise for $\left|
\omega \right| $ 
\mbox{$<$}
$30$ cm$^{-1}$ corresponding to a peak centered between $\pm $5 cm$^{-1}$.
This low frequency feature is asymmetrical around $\omega $ = 0, i.e., $%
\left| t^{-}(H)/t(0)\right| $ $>$ $\left| t^{+}(H)/t(0)\right| $. Its field
dependence is slightly super linear. At high frequencies ($\left| \omega
\right| $ $\geq $ 30 cm$^{-1}$) the transmission ratio approaches unity in
both modes such that $\left| t^{+}(H)/t(0)\right| $ 
\mbox{$>$}
1 and $\left| t^{-}(H)/t(0)\right| $ 
\mbox{$<$}
1, with a minimum at -40 cm$^{-1}$. This high frequency chiral response is
found to scale linearly with magnetic field for $H$ 
\mbox{$<$}
14$T$ $\ll $ $H_{c2}$\cite{Lihn} and has been interpreted in terms of the
tail of a free hole-like cyclotron resonance response\cite{Karrai}.

The overall shape of the transmission ratio spectrum in Fig.\ref{t9} is
simple which suggests that the underlying physics of vortex electrodynamics
may be very simple and elegant. Therefore it is interesting to convert the
transmission data into the conductivity function which is more useful for
gaining insight into the phenomena. We do this by means of a Kramers-Kronig
transformation technique. The most convenient quantity related to the
transmission experiment that satisfies the KKT conditions is $\ln (t(\omega
))$ $=$ $\ln \left| t(\omega )\right| +$ $i\,\arg [t(\omega )]$ in which $%
\arg [t(\omega )]$ is the phase of the complex amplitude $t(\omega )$.

In the zero field case, $\arg [t(\omega ,0)]$ can be obtained through KKT by
properly choosing an extrapolation function $t_{ext}(\omega )$ for $\left|
t(\omega ,0)\right| $, which preserves time reversal symmetry (even function
in $\omega $) and has correct asymptotic behavior at $\omega \rightarrow 0$ (%
$t_{ext}(\omega )\propto \omega $) and $\infty $ ($t_{ext}(\omega
)\rightarrow 1$). $t(\omega ,0)$ can then be converted into $\sigma (\omega
,0)$ since $Z_0$, $d$, and $n_{Si}$ are known. Detailed analyses show that
the film consists of $f_{s0}$ = 0.61 of superfluid condensate with the
London penetration depth $\lambda _0=$ 1850\AA \cite{background}. $Re[\sigma
(\omega ,0)]$ is shown in Fig.\ref{condpn}.

When a magnetic field is applied to the system, time reversal symmetry is
broken and it is then necessary to determine the proper response function
for both positive and negative frequencies. In this case, the two circularly
polarized modes are the canonical modes. We can extend the response function 
$g^{+}(\omega ,H)$ to the negative frequency range by 
\begin{equation}
g^{+}(\omega ,H)\equiv 
{g^{+}(\omega ,H)\quad ,\;when\;\omega >0\;\text{(eCP)} \atopwithdelims\{. g^{-}(-\omega ,H)^{*},\;when\;\omega <0\;\text{(hCP)}}
\label{circmap}
\end{equation}
where $g^{\pm }(\omega ,H)$ are the physical quantities measured for
positive frequencies in two circularly polarized modes. The KKT then is
extended to the magnetic field case, 
\begin{equation}
\arg [\frac{t^{+}(\omega ,H)}{t(\omega ,0)}]=\frac 1\pi P\int_{-\infty
}^\infty \frac{\ln \left| t^{+}(\omega ^{\prime },H)/t(\omega ^{\prime
},0)\right| }{\omega -\omega ^{\prime }}d\omega ^{\prime }.
\label{kk:realtran}
\end{equation}
By combining this with the zero field data from which $t(\omega ,0)$ is
extracted, we can obtain $\sigma ^{+}(\omega ,H)$ ($-\infty <\omega <\infty $%
) and therefore $\sigma ^{\pm }(\omega ,H)$ ($0<\omega <\infty $).

Fig.\ref{condpn} shows the magneto-conductivity $Re[\sigma ^{+}(\omega ,H)]$
obtained by KKT on the transmission curves in Fig.\ref{t9} with a $\ln
\left| t^{+}/t^{-}\right| \propto \omega ^3$ extrapolation scheme for $%
\left| \omega \right| <$ 5 cm$^{-1}$, which will be described later in this
letter. Apart from the residual metallic background which is present in $%
Re[\sigma (\omega ,0)]$\cite{background}, the conductivity function evolves
from its zero field London form ($\frac{ne^2}m\,f_{s0}\,i/\omega $) into a
form with a reduced London component and several finite frequency absorption
bands. The resulting conductivity can be well represented as a finite sum of
Lorentzian oscillators $\sigma _H(\omega ,H)=$ $ne^2/m$ $\sum_{i=1}^M$ $%
f_i\left/ \left( i(\omega -\omega _i)+\Gamma _i\right) \right. $ where $f_i$
represents the strength of the $i$th oscillator. Indeed, we found that {\it %
two} finite width oscillators ($M=2$) in addition to a reduced strength
London term ($\omega _0$=0, $\Gamma _0$=0) is sufficient to describe the
main features induced by vortex dynamics. The best fit with $M=2$ in $\sigma
_H(\omega ,H)$ is shown as the dotted line in Fig.\ref{condpn}. The first
oscillator is at low frequency ($\omega _1=$3.15 cm$^{-1}$, in eCP mode)
with $f_1$=0.14 (23\% of $f_{s0}$) and a width $\Gamma _1=$10 $cm^{-1}$\cite
{Parks} which is similar to the form of the G-R model\cite{GR}. The second
oscillator is centered at $\omega _2=$ -24 $cm^{-1}$ (in the hCP\ mode) with 
$f_2$= 0.11 (18\% of $f_{s0}$) and a width $\Gamma _2=$17 $cm^{-1}$, which
produces the optical activity observed at higher frequencies\cite{Vortex}.
The remaining oscillator strength gives $f_0=0.36$ (59\% of $f_{s0}$) since $%
f_0+f_1+f_2=f_{s0}$.

A sum rule on $t(\omega )$ follows from the superfluidity of the condensed
state. For pinned type II superconductors the superfluid condensate response
causes the low frequency conductivity to be dominated by the London
screening, $\sigma (\omega \rightarrow 0)\sim i/\omega +\pi \delta (\omega )$%
. Therefore $t^{+}(0,H)/t(0,0)$ is the ratio of the strength of two delta
functions which is a real number (for $H<H_{c2}$). The left hand side of Eq.(%
\ref{kk:realtran}) is zero (when $\omega \rightarrow 0$) which leads to the
relation 
\begin{equation}
\int_0^\infty \ln \left| t^{+}(\omega ^{\prime },H)/t^{-}(\omega ^{\prime
},H)\right| \frac{d\,\omega ^{\prime }}{\omega ^{\prime }}=0.
\label{supersum}
\end{equation}
This sum rule provides a strong constraint on the magneto-transmission data
which is useful for setting the extrapolations in the KKT analysis. It also
allows inferences to be made on the electrodynamics of the vortex system
outside the measurement range. Fig.\ref{sumrule} shows $\left| t^{+}(\omega
,H)/t^{-}(\omega ,H)\right| $ as a function of $\ln \left( \omega \right) $
and the integral weight of Eq.(\ref{supersum}) in the various regions.
Assuming that free electron behavior eventually dominates the free carrier
response at sufficiently high frequencies, then $\ln \left|
t^{+}/t^{-}\right| /\omega \sim \omega _c/\omega ^4\tau $ which indicates a
strongly convergent (superconvergent) sum rule. Indeed, extrapolation of the
broadband data by a simple cyclotron resonance model suggests that the
frequency range beyond 190 cm$^{-1}$ contributes only about 10\% to the
integral.

As shown in Fig.\ref{sumrule}, we have confirmed the observation of a sign
reversal of the optical activity $(T^{+}/T^{-}-1)$ in YBCO films (the same
batch of samples as in this letter.) at about 30 cm$^{-1}$, reported by Choi 
{\it et al.}\cite{Choi,Vortex}. Remarkably, this feature is a natural
consequence of Eq.(\ref{supersum}) since the weight given by the hybridized
hole cyclotron resonance\cite{Hsu,Karrai,Choi} at higher frequencies has to
be balanced by an electron-like Hall effect at lower frequencies.

Even though several different extrapolation schemes with reasonable physical
assumptions in the unmeasured regions ($\left| \omega \right| <$ 5.26 cm$%
^{-1}$ and $\left| \omega \right| >$ 200 cm$^{-1}$) give similar results and
preserves the two oscillator picture of the conductivity, the use of Eq.(\ref
{supersum}) helps to refine the analysis. A simple cubic spline
interpolation between $\pm $5.26 $cm^{-1}$ gives $\ln \left|
t^{+}/t^{-}\right| \sim \omega $, which results in $\sim $30\% excess weight
in the low frequency region compared to the high frequency region (double
dotted dash line in Fig.\ref{sumrule}). (Note that $1/\omega $ in Eq.(\ref
{supersum}) dramatically increases the weight of the low frequency optical
activity.) There are two possible resolutions of this discrepancy. The first
possibility is that there are other (electron-like) chiral resonances in
addition to the simple cyclotron resonance tail above 200 cm$^{-1}$ which
balances the low frequency weight. Simulations show that an additional 7\%
mid-IR chiral resonance at $\sim $500 cm$^{-1}$ will allow the sum rule to
be satisfied. However, this resonance would cause the optical activity to
change sign twice below 500 cm$^{-1}$ and there has been no evidence for
this in our measurements up to 200 cm$^{-1}$\cite{Choi,Vortex}$.$ The second
possibility is that $\ln \left| t^{+}/t^{-}\right| $ decreases faster than $%
\omega $ at low frequencies (below 5 cm$^{-1}$). By choosing an $\omega ^3$
extrapolation for $\ln \left| t^{+}/t^{-}\right| $ (which is an odd function
of $\omega $) we found that the sum rule can be satisfied (dashed line below
5 cm$^{-1}$). It turns out that the two-Lorentzian model mentioned above
with a {\it slightly} {\it electron-like} low frequency oscillator satisfies
the second criteria and is used as the extrapolation function for $|\omega
|< $ 5.26 cm$^{-1}$ in Fig.\ref{t9} (dotted line). This self-consistency
further convinces us that the two-Lorentzian model properly represents the
main response of the system.

We also note that the FIR data can be used to infer information about the
low frequency response of the system. $\ln \left| t^{+}/t^{-}\right| /\omega
\propto Re[\sigma _{xy}]$ at low frequencies where $Z_0d\sigma \gg n_{Si}+1$%
, $\sigma _{xx}\sim 1/i\omega $ and $\sigma _{xy}\ll \sigma _{xx}$.
Therefore $\ln \left| t^{+}/t^{-}\right| \propto \omega $ implies that the
low frequency limit of $Re[\sigma _{xy}]$ is a finite constant and $\rho
_{xy}\propto \omega ^2$. However if the second possibility holds true, an $%
\omega ^3$ dependence of $\ln \left| t^{+}/t^{-}\right| $ would imply $%
Re[\sigma _{xy}]\propto \omega ^2$ (dotted line in Fig. \ref{condxy}(b)) so
that $\rho _{xy}$ would go to zero as $\omega ^4$, corresponding to a very
strong suppression of the Hall effect. Therefore $\mu $wave studies of the
Hall effect in type II superconductor may prove interesting.

Fig.\ref{condxy}(a) shows $Im[\sigma _{xx}(\omega ,H)]/Im[\sigma (\omega
,0)] $ which describes the modification of the screening in the applied
field. This ratio approaches $0.55$ as $\omega \rightarrow 0$. In terms of
the G-R model\cite{GR} $\frac{Im[\sigma _{xx}(0,H)]}{Im[\sigma (0,0)]}%
=\kappa /(\kappa +\frac{c^2}{4\pi \lambda _0^2}\frac{H\phi _0}{c^2})$ which
gives $\kappa \simeq 5.3\times 10^5$ N/m$^2$. This value of $\kappa $ is
consistent with diverse $\mu $wave measurements on films and single crystals%
\cite{Golo,Parks,Revenaz}, suggesting an intrinsic origin for $\kappa $.

There are other smaller structures (%
\mbox{$>$}
50 cm$^{-1}$) in the conductivity that correspond to only a few percent (or
less) of the total oscillator strength. Many of these features are
reproducible, scale in amplitude with $H$ and are found to correlate with
the density of 45$^{\circ }$ misaligned grains in this sample. They have
been discussed elsewhere in terms of vortex core excitations\cite{Lihn}.
Also we note that $(T^{+}/T^{-}-1)$ at low frequencies (7 $\sim $ 15 cm$%
^{-1} $) decreases linearly with temperature and changes its sign at $\sim $%
20 K. This behavior is consistent with cyclotron resonance of thermally
excited quasiparticles as has been reported by Spielman {\it et al.}\cite
{Spielman}. This observation shows that the low temperature optical activity
is dominated by vortex dynamics.

The nonzero center frequency of the low frequency oscillator is a
consequence of the low frequency extrapolation as determined from the
superconvergent sum rule. Within the G-R type models it implies a Magnus
force, $ne\left( v_s-\beta v_L\right) \times \phi _0$ where $\beta \approx
-0.08$\cite{Newmodel}. The negative $\beta $ is consistent with observation
of a reversal of dc\ Hall effect in the flux flow regime\cite{Dorsey,Hagen}.
No evidence for optical activity has been reported in the low temperature
vortex response at $\mu $wave frequencies. This may be related to the strong
suppression of the Hall resistance discussed above. In the G-R model optical
activity requires a Magnus force, but the zero vortex mass keeps the
resonance at low frequency. Hsu's model contains a Magnus force, core
excitations and, implicitly, a vortex mass and it produces two finite
frequency chiral resonances. In the absence of pinning the Hsu model
predicts that the core resonance is silent and the response of the
superconductor is simple cyclotron resonance as expected from Kohn's theorem%
\cite{Drew}. Pinning hybridizes the cyclotron resonance and vortex core
resonance with the pinning resonance. The result is a strong hybridized
pinning resonance at a negative frequency, which gives optical activity and
a weak hybridized vortex core resonance at a positive frequency.

The observed response is seen to contain the features of both theories, but
is inconsistent with either one alone. It appears that there is a massless
response of the vortices that gives rise to the low frequency oscillator in
addition to a finite vortex mass response that gives rise to the pinning
resonance and the weak vortex core resonance in the eCP mode\cite{Lihn,Choi}%
. These observations suggest a model in which the vortex is considered as a
composite object. The observed responses can then arise from the massless
response of the vortex currents outside the core and the inertial response
of the core. The observed universal value of $\kappa $ then is a natural
consequence of this model due to the pinning of the vortex current pattern
to the vortex core. A paper describing this model will be presented elsewhere%
\cite{Newmodel}.

The authors acknowledge helpful discussions with V. Yakovenko. This work is
supported by National Science Foundation under Grant No. DMR9223217. The
work at Advanced Fuel Research was supported by Contract No. DOE-SBIR
(DE-FG01-90ER81084 and DE-FG05-93ER81507).

\begin{figure}[tbp]
\caption{The transmission amplitude ratio $\left| t^{+/-}(\omega
,H)/t(\omega ,0)\right| $ as a function of frequency $\omega $ for a YBCO
thin film at $H$=9$T$ and 4 K for the circularly polarized light. The solid
curve represents the FTS data from 30 cm$^{-1}$ to 200 cm$^{-1}$ and the
triangular points represent the data from the FIR laser lines. The region
between 125$\sim $140 cm$^{-1}$ corresponds to a quartz phonon and the half
waveplate condition of the polarizer. The dotted line is the $\ln \left|
t^{+}/t^{-}\right| \sim \omega ^3$ extrapolation described in text. Inset :
Zero field transmission amplitude $\left| t(\omega ,0)\right| $ vs.
frequency $\omega $ at 4$^{\circ }$K (the solid line). The dotted line is
the extrapolation function $t_{ext}(\omega )$.}
\label{t9}
\end{figure}

\begin{figure}[tbp]
\caption{The magneto-conductivity $Re[\sigma^{+}(\omega ,H)]$ obtained from
the Kramers-Kronig analysis. The change induced by the magnetic field can be
described as the sum (dotted line) of a low frequency oscillator of width $%
\Gamma _1=$ 10 cm$^{-1} $ at $\omega _1=$ +3.15 cm$^{-1}$ (single dotted
dash line) and an oscillator of width $\Gamma _2=$ 17 cm$^{-1}$ at $\omega
_2=$ -24 cm$^{-1}$ (double dotted dash line). The small dotted line is $%
Re[\sigma (\omega ,0)]$.}
\label{condpn}
\end{figure}

\begin{figure}[tbp]
\caption{$\left| t^{+}(\omega ,H)/t^{-}(\omega ,H)\right| $ is plotted as a
function of $\ln \left( \omega \right) $. The spectral weights of different
frequency regions to the sum rule of Eq.(\ref{supersum}) are indicated. The
dashed line between laser data points is a guide for the eye. The double
dotted dash line is the simple cubic spline extrapolation which behaves as $%
\ln \left| t^{+}/t^{-}\right| \sim \omega $ below 5 $cm^{-1}$. The dashed
line is an $\omega ^3$ extrapolation for $\ln \left| t^{+}/t^{-}\right| $
which satisfies the sum rule.}
\label{sumrule}
\end{figure}

\begin{figure}[tbp]
\caption{The magneto-conductivity shown in the Cartesian coordinates, $%
\sigma _{xx}$ and $\sigma _{xy}$. Dotted lines are the two-Lorentzian model
fits described in Fig.\ref{condpn}. (a) $Im[\sigma _{xx}(\omega
,H)]/Im[\sigma (\omega ,0)]$ represents the reduction of screening in the
applied field. (b) $Re[\sigma _{xy}(\omega ,H)]$ and (c) $Im[\sigma
_{xy}(\omega ,H)]$ show a resonance at 24 $cm^{-1}$.}
\label{condxy}
\end{figure}

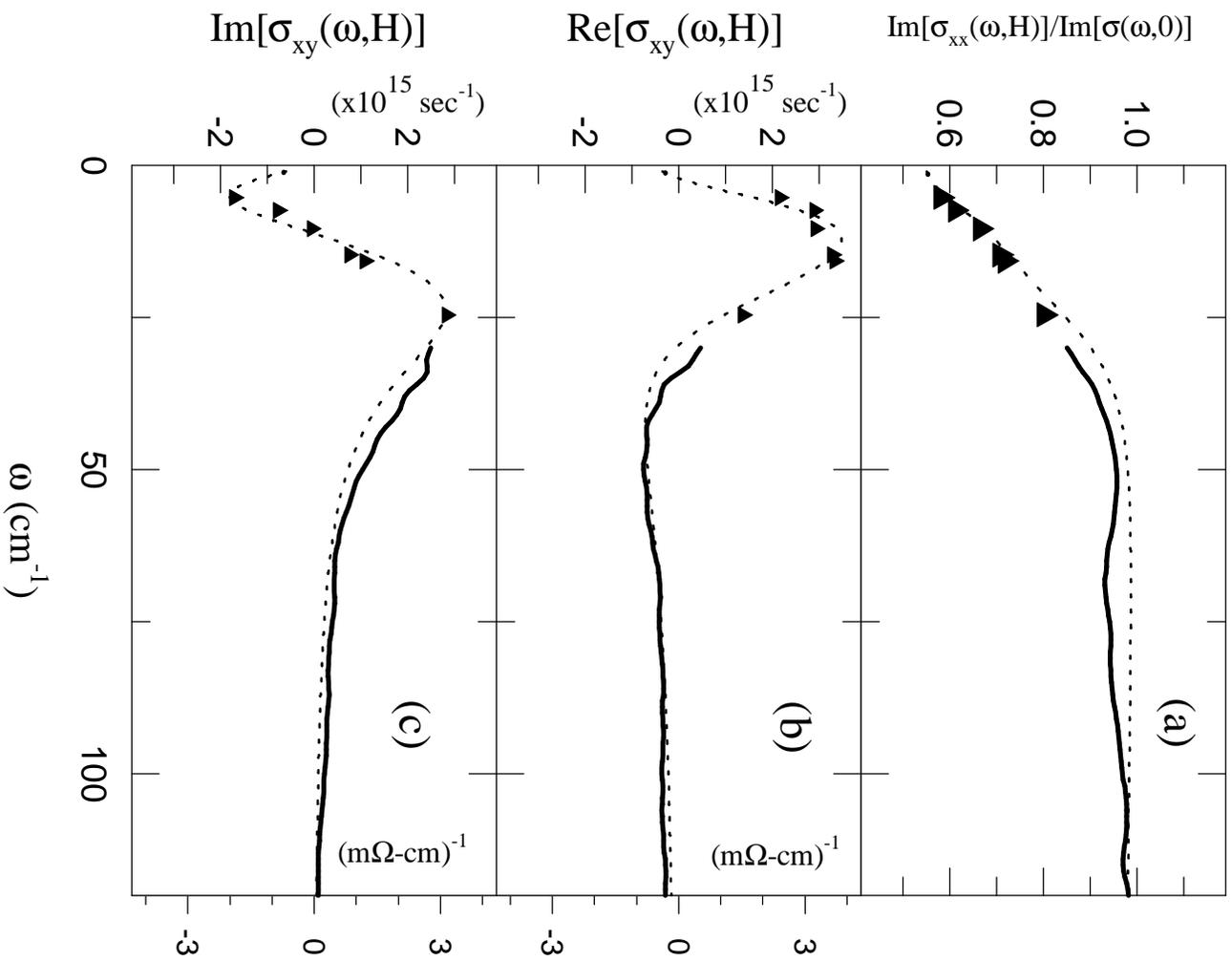
\end{document}